\documentclass[12pt]{article}
\usepackage{latexsym}
\usepackage{a4,amssymb, amsthm, amsmath}

\newtheorem{theorem}{Theorem}
\newtheorem{lemma}[theorem]{Lemma}
\newtheorem{proposition}[theorem]{Proposition}
\newtheorem{definition}[theorem]{Definition}

\newtheorem{remark}[theorem]{Remark}

\begin{document}

\title{Fixed Energy R-separation for Schr\"odinger equation}
\author{Claudia Chanu \& Giovanni Rastelli}
\date{ }

\maketitle

\begin{abstract}
We extend the classical approach of the $R$-separation of the Laplace equation $\Delta \psi=0$ (as a null eigenvalue problem) to the general steady state Schr\"odinger equation including cases with a scalar potential $V$  and the energy
is a fixed constant. 
\end{abstract}

Keywords: {Variable separation; Schr\"odinger equation; Laplace equation; R-separation.}



\section{\large Introduction}

Let us consider the Schr\"odin\-ger equation
 \begin{equation} \label{sch}
-\frac {\hbar^2} 2 \Delta \psi +(V-E)\psi=0,
  \end{equation}
on a Riemannian manifold $Q$ with metric tensor $\mathbf G$ of any signature. 
According to Moon and Spencer \cite{MS52, MS52b, MSb}, if the ansatz 
$$
\psi=R\textstyle\prod_i\phi_i(q^i),
$$
where $R$ is a suitable function of the coordinates
and each $\phi_i$ is a function of the corresponding coordinate $q^i$ only, 
permits to split equation (\ref{sch}) into $n$ separated ODE's, then this equation is said to be {\em R-separable}.
%
We are interested in the following question: under which conditions on the metric $\mathbf G$, the potential $V$ and the value of the energy $E$ is this ansatz successful?

Usually, in the literature \cite{MS52, MS52b, MSb,KMto, KMoRsep, KMesRsep, KM4dR}, authors require that $R$-separable solutions  for Schr\"odin\-ger or Helmholtz equations exist for all possible eigenvalues $E$.
In the present paper we study the more general theory in which $R$-separation occurs for a single eigenvalue $E$, 
as in the Laplace equation, where the only eigenvalue is $E=0$. We call this theory {\em fixed energy R-separation} ({\em FER-separation}). 
As a result, we shall see that FER-separation enlarges the number of classes of separable coordinates. 
First of all we analyse the definition of FER-separation, showing that the functions $\phi_i$ must depend on a suitable number of parameters $(c_\alpha)$ in order to satisfy a completeness condition (Definition \ref{d:rsep}). 
According to this definition,
we will provide a criterion for testing if FER-separation occurs in a given coordinate system $(q^i)$ (Theorem \ref{t:res}): we shall see that a necessary condition is that coordinates be orthogonal and 
{\em conformal separable} (as defined in \cite{hje}).
Then, we give necessary  intrinsic conditions for the existence of FER-separable coordinates based on conformal Killing tensors (Proposition \ref{p:intr}). 
Furthermore, ODEs whose integrations provide $R$-separated solutions of (\ref{sch}) are deduced and the associated second order commuting operators are determined. 
In such a way, we see that our definition of $R$-separation is equivalent, when restricted to the Laplace equation, to the 
definition of orthogonal separable coordinates given by Kalnins and Miller \cite{KMLapl} involving a set of operators constructed from a St\"ackel matrix. Indeed, the R-separation of Laplace equation is characterized in \cite{KMLapl} by means of {\em conformal symmetry operators}.
We remark that in our approach, both the orthogonality and the existence of a related St\"ackel matrix are proved and not assumed.

Also the intermediate results, the ``tools" used for proving  Theorem \ref{t:res}, appear to be quite interesting by their own: they are (i) a geometric interpretation of the additive separation of variables in a single null PDE of any order, (ii) three criteria for additive separation of such a PDE, which extend those given for the null Hamilton-Jacobi equation  in \cite{hje}, both exposed in Section \ref{s:HJE}. 
This geometrical approach is quite general and easily  adapted for dealing with fixed energy separation and FER-separation of other PDE's of physical interest, like time-dependent Schr\"odinger equation or  Schr\"odinger equation with a vector potential.  
For instance, as a straightforward application of the method, we get that, on an Einstein manifold, ordinary separation for a single value of $E$ of (\ref{sch}) only occurs in the coordinates allowing separation for all admissible values of $E$ (this is well-known for the  Laplace equation, which on Einstein manifolds is separable in the same coordinates separating Helmholtz equation). In other words, studying the fixed energy case of ordinary separation is not so interesting, since in the most common cases we do not get any new coordinate system, while for  FER-separation  the class of possible coordinates is definitively bigger.
This is explain the importance of FER-separation, whose theory, even if applied to Laplace equation till the 19-th century, needs a precise mathematical foundation.

\section{\large From multiplicative R-separation to additive separation} \setcounter{equation}{0}
The link between $R$-separation of (\ref{sch}) and additive separation of a suitable PDE depending on $\mathbf G$,
$V$, $R$, and $E$ is obtained via the usual (see for instance \cite{KMto}) substitutions,
\begin{equation}\label{u}
u=\ln \phi
 \end{equation}
\begin{equation}\label{rsol}
\psi=R\phi.
\end{equation}
 We recall that the Laplace-Beltrami operator $\Delta=g^{ij}\nabla_i\nabla_j$ on a  $n$-dimensional Riemannian or pseudo-Riemannian manifold $(Q,\mathbf G)$ with contravariant metric tensor $\mathbf G =(g^{ij})$ in local coordinates has the form
\begin{equation} \label{lap}
\Delta \psi=g^{ij}\partial^2_{ij}\psi-\Gamma^h \partial_h\psi.
\end{equation}
where  $\Gamma^h$ are the {\em contracted Christoffel symbols}, defined as  
 \begin{equation}\label{ccs}
\Gamma^h=g^{ij}\Gamma^h_{ij}, \qquad \Gamma_k=g_{kh}\Gamma^h.
  \end{equation}
  
\begin{proposition}\label{p:equi}
There is a one to one correspondence between the solutions of $(\ref{sch})$ 
 of the form
$
\psi=R\prod_i\phi_i(q^i)
$
and the additively separated solutions $u=\ln \phi$ of 
  \begin{equation} \label{nbhj}
g^{ij}u_iu_j+g^{ii}u_{ii}-\hat \Gamma^i u_i+\textstyle\frac 2{\hbar^2}E-U=0,
  \end{equation}
where $u_i=\partial_i u$, $u_{ii}=\partial^2_{ii}u$, $U$  is the modified potential
 \begin{equation} \label{nbarU}
U=-\left(\frac{\Delta R}R-\frac 2{\hbar^2}V\right),
  \end{equation}
and  
\begin{equation} \label{hatG}
\hat \Gamma^i=g^{ij}(\Gamma_j-2\partial_j\ln R).
\end{equation} 
\end{proposition}  
  
\begin{proof} 
By inserting (\ref{rsol}) in (\ref{sch}) for a real function $R\neq 0$ on $Q$,
we see that (\ref{rsol}
is a solution of 
$(\ref{sch})$ if and only if
$\phi$ satisfies 
  \begin{equation} \label{bar1}
R\Delta \phi+2\nabla R \cdot \nabla \phi+\left(\textstyle\frac 2{\hbar^2}R(E-V)+\Delta R\right)\phi=0.
  \end{equation}
which is equivalent to 
 \begin{equation} \label{nbar1}
\Delta \phi+2\nabla\ln R\cdot \nabla \phi+\left(\textstyle\frac 2{\hbar^2}E-U\right)\phi=0,
  \end{equation}    
where $U$ is given by (\ref{nbarU}). 
Then, let us perform the substitution (\ref{u}) and let us
denote the partial derivatives of $u$ by
$
u_i=\partial _iu,$ 
$u_{ij}=\partial ^2_{ij}u,$ $u_{ijk}=\partial ^3_{ijk}u,$ etc.
Since $u$ is additively separated, we get 
$
u_{ij}=\delta ^i_ju_{ii}.
$
Moreover, by (\ref{u}) we have
  \begin{equation}\label{der}
\partial _i\phi=u_i\phi,\qquad
\partial^2_{ii}\phi =(u_{ii}+u_i^2)\phi,\qquad
\partial _{ij}\phi=u_iu_j\phi \quad (i\ne j).
  \end{equation}
Therefore, 
by inserting (\ref{der}) in (\ref{nbar1}), due to (\ref{ccs}), we get 
$$(g^{ij}u_iu_j+g^{ii}u_{ii}-\hat \Gamma^i u_i+\textstyle\frac 2{\hbar^2}E-U)\phi=0.$$
Hence, since $\phi\neq 0$, $\phi$ is a multiplicatively separated solution of (\ref{nbar1}) if
and only if  $u$ is a separated solutions of (\ref{nbhj})  . 
\end{proof}

\begin{remark} \rm
Although Eqs. (\ref{bar1}) and (\ref{nbar1}) admit the same solutions for any choice of $R\neq 0$, 
they are deeply different from the point of view of separability:
equation (\ref{nbar1}) is treated in a natural way as an eigenvalue problem and its separation for all the eigenvalues $\frac{2}{\hbar^2}E$ leads to $R$-separation theory as developed 
by Moon and Spencer \cite{MS52,MS52b} and by Kalnins and Miller \cite{KMesRsep,KM4dR,KMoRsep,KMnoRsep}
On the contrary, in (\ref{bar1}) the energy constant $E$ is multiplied by the function $R$ and $E$ cannot be considered as a differential operator eigenvalue: in (\ref{bar1}) $E$ is  a fixed parameter incorporated into the differential operator and the solutions are the eigenfunctions of the null eigenvalue only. Its separation must be treated 
in a way similar to the separation for Laplace equation \cite{MS52,KMLapl}. 
The difference between these two view points is subtle but relevant (see also \cite{Marmo} about the  correspondence between the quantum-mechanical description of hydrogen atom and the harmonic oscillators).  
Of course,  also Eq.\ (\ref{nbar1}) can be read as a null eigenvalue problem by including the constant $E$ into the potential function $U_E=\frac 2{\hbar^2}E-U$.
We adopt this last interpretation of (\ref{nbar1}) to define Fixed Energy R-separation (FER-separation) in Section
\ref{s:R-sep}. 
In this approach the function $R$ plays the same role of the potential $V$ and the metric tensor $(g^{ij})$: 
Eq.\ (\ref{nbar1}) is a PDE defined by $\mathbf G$, $V$, $R$ and $E$. By imposing that (\ref{nbar1})
admits a complete family of separated solutions $\phi$, in a meaning we will make precise later,
we determine differential conditions on $\mathbf G$, $V$, $R$ and $E$.
\end{remark}

\begin{remark} \rm
The change of unknown function (\ref{u}) relates multiplicative and additive separation of any PDE,
both in the case in which we are dealing with a single null PDE or with an eigenvalue problem.
In the resulting equation (\ref{nbhj}) the parameter $E$ admits two interpretations as in equation (\ref{nbar1}):
it can be considered as an internal parameter of the equation (FER-separation) or as an integration constant (classical R-separation).
\end{remark}


\section{\large Geometry of the additive separation for a null PDE of order $l$}\label{s:HJE}
\setcounter{equation}{0}
We provide the geometric framework and the geometric characterization of the
separation of variables (SOV) for a null PDE of arbitrary order $l$, extending the results
of \cite{hje} (concerning the Hamilton-Jacobi case only) and of \cite{schr1} (dealing with ordinary separation only).
The main tool is the theory of separable connections as introduced by Benenti \cite{BeSepCo}. 
As in the case of the Hamilton-Jacobi equation with a fixed value of the energy, two different 
(but in a sense equivalent)
definitions of separable solutions are given. Each of them is useful to prove different criteria of existence of a separable solution in a given coordinate system. 


\subsection{Separable connections}\label{ex2.1}
Let $M=Q\times Z\to Q$ be the trivial vectorial bundle with fiber $Z$ over $Q$, where $Q$ is a $n$-dimensional manifold
with coordinates $(q^i)$ ($i=1,\ldots,n$) and $Z$ is a $N$-dimensional vector space (over $\mathbb R$ or $\mathbb C$)
with coordinates $(z^A)$ ($A=1,\ldots, N)$.

We call  {\em connection} on $M$ a regular distribution (in the Frobenius sense) $\mathcal C$ on $M$ of rank $n$ and transversal to the fibers: $\mathcal C:x\mapsto \mathcal C_x\subset T_xM$ with $\dim \mathcal C_x=n$ and
$\mathcal C_x\cap T_xZ=\{\mathbf 0\}$.

A vector field $X\in \mathcal X(M)$ is horizontal if $X(M)\subset \mathcal C(M)$ i.e., if $X(x)\in \mathcal C_x$.
Locally a connection $\mathcal C$ is spanned by the $n$ horizontal vector fields, called the {\em generators} of $\mathcal C$
$$
D_i=\frac{\partial}{\partial q^i}+ C_{iA}\frac{\partial}{\partial z^A},
$$
where $C_{iA}$ are functions on $M$ called {\em coefficients} of the connection. 
A connection is {\em integrable} (in the Frobenius sense) if for any point $P\in M$ there exists a unique integral manifold (i.e., a $n$-dimensional immersed submanifold tangent to each $D_i$)
passing through $P$.
By the Frobenius theorem and  the particular form of the generators, we get that

\begin{proposition}{\it A connection generated by the vector fields $(D_i)$ is integrable (in the Frobenius sense) if and only if $[D_i,D_j]=0$.}
\end{proposition}

If $\mathcal C$ is integrable, then there exists a foliation $\mathcal L$ of $M$ made of integrable $n$-dimensional manifolds of $\mathcal C$. 
Since $\mathcal C$ is a transversal distribution, the integral manifolds
 are locally described by $z_A=F_A(q^i,c_B)$ where $(c_B)$  are real parameters ($B=1,\ldots,N$)satisfying 
the completeness condition
$$
\det\left[\frac{\partial z_A}{\partial c_B}\right]\neq 0.
$$
The functions $F_A$ are solutions of the first-order differential system of PDE on $Q$
$$
\frac{\partial z_A}{\partial q^i}=C_{iA}(q^i,z_A).
$$
It follows that

\begin{proposition}
 The connection generated by the $D_i$ is integrable if and only if the corresponding system of PDE is completely integrable.
\end{proposition}

Let $S$ be a $N-1$-dimensional submanifold of $M$.
We say that a distribution $\mathcal C$ is {\em reducible} on $S$  if when it is restricted to the points of $S$ it gives rise to a distribution $\mathcal C_0$ on $S$. We have that  

\begin{proposition}
 A distribution is reducible to a submanifold $S$ if and only if its generators $D_i$ are tangent to  $S$.
\end{proposition}

If $F$ is a function on $M$ such that equation $F=0$ implicitly defines the submanifold $S$,
then $\mathcal C$ is reducible to $F=0$
if and only if $D_iF|_{S}=0$ or, equivalently,  $D_iF=\Lambda F$ for a suitable function $\Lambda$ on $M$.

\begin{remark} \rm
If a connection $\mathcal C$ is integrable and reducible as a distribution to $S$, then the foliation $\mathcal L=(L_{c_A})$ on $M$ made of integral manifolds of $\mathcal C$
can be restricted to a foliation $\mathcal L_S$ on $S$ made of integral manifolds of $\mathcal C$, whose labels $(c_A)$ belong to a suitable 
$N-1$-dimensional submanifold $\mathcal P$ of $\mathbb R^N$. Up to a new parametrization of the $(c_A)$, we can assume without loss of generality that $\mathcal P$ is defined by $c_N=0$ and $\mathcal L_S$ is described by the $N-1$ parameters $(c_\alpha)=(c_1,\ldots,c_{N-1})$.
\end{remark}

\subsection{Equivalent definitions of additive SOV for a null PDE}
Inspired by the approach to the additive separation of Kalnins and Miller exposed in \cite{KMto} for PDE
of the kind $\mathcal H=\mathrm{const.}$, we extend the theory in a geometrical way to a 
single PDE of order $l$ on $Q$
\begin{equation}\label{eq}
\mathcal H(q^1,\ldots, q^n, u,u_i,u_{ij},\ldots,u_{ij\ldots h})=0, 
\end{equation}
where $u_i=\partial_i u$, $u_{ij}=\partial^2_{ij}u$ $\ldots$ are the derivatives (up to order $l$) of the unknown function $u=u(q^i)$, with respect to the coordinates, with $\partial_i=\partial/\partial q^i$.
An (additive) {\em separated solution} of (\ref{eq}) is a function of the form
$$
u=\sum_{i=1}^n S_i(q^i),
$$
where $S_i$ depends on the single variable $q^i$
satisfying (\ref{eq}).
If we look for additive separated solutions only, Eq.\ (\ref{eq}) reduces to
$$
\mathcal H(q^1,\ldots, q^n, u,u_i,u^{(2)}_{i},\ldots,u_i^{(l)})=0.
$$
where $u^{(2)}_i=u_{ii}$ and $u_{ij}=0$ for $i\neq j$.

\begin{definition} \rm
Given a PDE $(\ref{eq})$ of order $l$, the  function $u^I$ is a {\em complete internal separated solution}
(internal solution) if: 
i) $u^I$ is of the form
$$
u^I=\sum_{i=1}^n S_i(q^i,c_\alpha), \qquad \alpha=1,\ldots, nl;
$$
ii) for all $(c_\alpha)$ in a suitable open set of $\mathbb R^{nl}$ $u^I$ is a separated solution of $(\ref{eq})$ i.e.
$$\mathcal H(q^1,\ldots, q^n, u,u^I_i,u^I_{ii},\ldots,u^I_{ii\ldots i})=0;$$  
iii) the following completeness condition holds
$$
\mathrm{rank}\left[\frac{\partial u^I}{\partial c_\alpha}  \; \bigg | \;  \frac{\partial u_i^I}{\partial c_\alpha}
 \; \bigg | \; \frac{\partial u_{ii}^I}{\partial c_\alpha}  \; \bigg | \;  \ldots  \; \bigg | \;  \frac{\partial u_{i\ldots i}^I}{\partial c_\alpha}\right]=nl.
$$
\end{definition}

\begin{definition} \rm
Given the PDE $(\ref{eq})$ of order $l$, the  function $u^E$ is a {\em complete extended separated solution} 
(extended solution) if:  
i) $u^E$ is of the form
$$
u^E=\sum_{i=1}^n S_i(q^i,c_a), \qquad a=1,\ldots, nl+1,
$$
ii)  $u^E$ satisfies $(\ref{eq})$, that is 
$\mathcal H(q^1,\ldots, q^n, u^E,u^E_i,u^E_{ii},\ldots,u^E_{ii\ldots i})=0,$ 
for any $(c_a)$ belonging to a suitable $nl$-dimensional submanifold of  $\mathbb R^{nl+1}$ or, up to a transformation of the $(c_a)$, for $c_{nl+1}=0$;
iii) $u^E$ satisfies the following completeness condition for all admissible values of $q^i$ and $c_a$:
$$
\det \left[\frac{\partial u^E}{\partial c_a}  \; \bigg | \;  \frac{\partial u_i^E}{\partial c_a}
 \; \bigg | \; \frac{\partial u_{ii}^E}{\partial c_a}  \; \bigg | \;  \ldots  \; \bigg | \;  \frac{\partial u_{i\ldots i}^E}{\partial c_a}\right] \neq 0.
$$
\end{definition}

\begin{remark} \rm
An internal complete solution defines a foliation of the submanifold $\mathcal H=0$ in $n$-dimensional leaves transversal to the fibers of the bundle $M$, via equations 
$$
u=u^I, \quad u_i=\partial_i u^I, \quad u_i^{(2)}=\partial^2_{ii} u^I,\quad \ldots  \quad u_i^{(l)}=\partial^l_{i\ldots i} u^I.
$$
Each submanifold is parametrized by the value of the $nl$ parameters $(c_\alpha)$.
Conversely,  an extended solution defines a foliation of an open subset of $M$ in $n$-dimensional leaves
via equations 
$$
u=u^E, \quad u_i=\partial_i u^E, \quad u_i^{(2)}=\partial^2_{ii} u^E, \quad \ldots \quad u_i^{(l)}=\partial^l_{i\ldots i} u^E.
$$
Each submanifold is parametrized by the value of the $nl$ parameters $(c_a)$.
The foliation is compatible with the submanifold $\mathcal H=0$ in the sense that it is reducible to a foliation
of $\mathcal H=0$.
\end{remark}

\begin{proposition}{\it
In the coordinates $(q^i)$ on $Q$, there exists an internal solution of the PDE $(\ref{eq})$
if and only if there exists an extended solution of $(\ref{eq})$ in the $(q^i)$ .
}
\end{proposition}

\begin{proof}
Let $u^E(q^i,c_a)$ be an extended solution of (\ref{eq}). Then, up to a reparametrisation of the $(c_a)$,
we can assume that for  all $(c_a)$ with $c_{nl+1}=0$ the function $u^E$ is a solution of (\ref{eq}).
Hence, $u^I(q^i,c_\alpha)=u^E(q^i, c_1,\ldots, c_{nl},0)$ is an internal solution for (\ref{eq}) in the same coordinates.
Conversely, let $u^I(q^i,c_\alpha)$ be an internal solution of (\ref{eq}).
Let $C_m$ be the first column in the matrix
$$
M^I=\left[\frac{\partial u^I}{\partial c_\alpha}  \; \bigg | \;  \frac{\partial u_i^I}{\partial c_\alpha}
 \; \bigg | \; \frac{\partial u_{ii}^I}{\partial c_\alpha}  \; \bigg | \;  \ldots  \; \bigg | \;  \frac{\partial u_{i\ldots i}^I}{\partial c_\alpha}\right]
 $$
such that the $nl$-th order minor $M^I_m$ obtained by eliminating this column is different from zero.
Such an index $m$ exists because of the completeness condition. We consider the two cases
$m=1$ and $m>1$ separately. If $m=1$, then $u^E=u^I+c_{nl+1}$ is a solution of (\ref{eq}) for all
$(c_a)$ such that $c_{nl+1}=0$; moreover, we have that 
$$
M^E=\left[\textstyle\frac{\partial u^E}{\partial c_a}  \; \bigg | \;  \textstyle\frac{\partial u_i^E}{\partial c_a}
 \; \bigg | \;  \ldots  \; \bigg | \;  \textstyle\frac{\partial u_{i\ldots i}^E}{\partial c_a}\right]=
\left[\begin{array}{c} \frac{\partial u^I}{\partial c_\alpha} \\ 1\end{array} \; \bigg | \; 
\begin{array}{c} \frac{\partial u_i^I}{\partial c_\alpha} \\ 0\end{array}
 \; \bigg | \; 
\begin{array}{c} \frac{\partial u_{ii}^I}{\partial c_\alpha} \\ 0\end{array} \; \bigg | \;  \ldots  \; \bigg | \;  \begin{array}{c} \frac{\partial u_{i\ldots i}^I}{\partial c_\alpha} \\ 0\end{array} \right] 
$$
and the completeness condition holds.
If $m>1$,  there exist unique $q,r\in \mathbb Z$ such that $m-2=qn+r$ with $0\leq r<n$. Thus,
 the eliminated column contains the partial derivatives w.r.t.  $(c_\alpha)$
 of 
 $\frac{\partial^ku^I}{(\partial q^j)^k}$ with $k=q+1$ and $j=r+1$.
Then, $$u^E=u^I+c_{nl+1}\frac{(q^{j})^k}{k!}$$ 
is a solution of (\ref{eq}) for all
$(c_a)$ such that $c_{nl+1}=0$. Moreover, the rank of
$$
M^E=\left[\frac{\partial u^E}{\partial c_a}  \; \bigg | \;  \frac{\partial u_i^E}{\partial c_a}
 \; \bigg | \; \frac{\partial u_{ii}^E}{\partial c_a}  \; \bigg | \;  \ldots  \; \bigg | \;  \frac{\partial u_{i\ldots i}^E}{\partial c_a}\right]
$$
is maximal.
Indeed, in the developing of $\det M^E$ along the last row, the first $m-1$ addenda are null 
since  $M^I_h=0$ for $h<m$, as well as those from the $(m+1)$-th to the last, because the elements of the last row of $M^E$ with column index  greater than $m$ are all null. Hence, being 
 $\det M^E=(M^E)^{nl+1}_{m} M^I_m=1\cdot M^I_m\neq 0$,
the completeness condition holds.
\end{proof}

\begin{definition} \rm
We say that the null equation $\mathcal H=0$ is {\em separable} in the coordinates $(q^i)$
 if it admits an extended solution or, equivalently, an internal one. 
\end{definition}

\begin{remark} \rm
In comparison with the geometric framework introduced in \cite{hje} to deal with a null Hamilton-Jacobi equation, the
present case presents several differences. First of all, it is not possible to define complete internal or extended solutions without assuming that they are separated; moreover there is not a unique formula transforming an internal into an extended solution. This is a consequence 
of the fact that the coordinates in the fiber are not all of the same kind, but correspond to derivatives of $u$ of different orders. 
\end{remark}

\subsection{Criteria for additive separation of a null equation}

Let us consider fibered coordinates $(q^i,z_A)$ on  $M=Q\times Z\to Q$,
where $(z_A)=(u,u_i,u_i^{(2)}, \ldots,u_i^{(l)})$, with $i=1,\ldots,n$, $A=1,\ldots,nl+1$. 
Let 
$\mathcal H$ be a smooth function on $M$ such that $\mathcal H=0$ 
is a $nl$-dimensional submanifold of $M$.
We apply to this case the equivalence of integrability of distributions on vector bundles and complete integrability of normal first-order differential systems, recalled in section \ref{ex2.1}.
The following propositions provide necessary and sufficient conditions for the existence of an extended and an internal solution, respectively (see also \cite{hje} and references therein). 


\begin{proposition} \label{critE}
In given coordinates $(q^i)$ on $Q$, the following statements are equivalent:
$(i)$ there exists an extended solution $u^E$ of $\mathcal H=0$;
$(ii)$ there exist $n$ functions $R_i$ on $M$ such that the distribution $\Delta$ generated by
  \begin{equation} \label{gene}
D_i=\partial_i+u_i\partial_u+u^{(2)}_{i}\partial/\partial u_i+\ldots+u_i^{(l)}\partial/\partial u_i^{(l-1)}+R_i\partial/\partial u_i^{(l)}
\end{equation}
is integrable and reducible on the submanifold $\mathcal H=0$, that is 
\begin{equation} \label{condiz}
\left\{
\begin{array}{l}
D_iR_j=0, \qquad \forall \, j\neq i \\  
D_i\mathcal H|_{\mathcal H=0}=0.
\end{array}
\right.
\end{equation}
\end{proposition}

\begin{proof}
$(i)\Rightarrow(ii)$:
Let us assume that a solution $u^E=\sum_iS_i(q^i,c_1,\ldots, c_N)$ of $\mathcal H=0$ exists. By the completeness condition, $(q^i,c_1,\ldots, c_N)$ are (non fibered) coordinates on $M$ and the  $n$ functions on $M$ 
  $$
R_i=\frac{d^{l+1}}{(dq^i)^{l+1}}S_i(q^i,c_1\,\ldots, c_N)
  $$ 
are well-defined. Going back the original coordinates we get functions 
$$R_i=R_i(q^i,u,u_i,u_{i}^{(2)},\ldots u_{i}^{(l)}),$$
and by them we construct the vector fields $D_i$ (\ref{gene}).
The distribution $\Delta$ generated by the fields $D_i$ is integrable because the corresponding first order system on $M$ 
  \begin{equation} \label{sys}
  \left\{
\begin{array}{l}
\partial_i u=u_i\\
\partial_i u_j=\delta_{ij} u_i^{(2)}\\
\ldots \\
\partial_i u_j^{(l)}=\delta_{ij} R_j
\end{array}
\right.
  \end{equation}
  is completely integrable.
Indeed,
$$
 u=u^E= \textstyle\sum_iS_i(q^i,c_A), \quad
 u_i=\textstyle\frac{d}{dq^i} S_i(q^i,c_A)\quad
\ldots \quad
u_i^{(l)}=\textstyle\frac{d^l}{(dq^i)^l} S_i(q^i,c_A)
  $$
is a complete solution of the PDEs system, which describes a foliation $\mathcal L_{(c_A)}=(L_{c_A})$
made of integral manifolds  of $\Delta$.
Moreover, since $u^E$ is a solution of  $\mathcal H=0$ for any $(c_A)$ with $c_{nl+1}=0$,
 we have $\mathcal H(q^i,u^E,\ldots)=0$, that is every point of $L_{c_A}$ belongs to  $\mathcal H=0$.
 Thus, the generators $D_i$ (by construction tangent to every leaf of $\mathcal L_{c_A}$) are tangent to 
 $\mathcal H=0$. Hence, $\Delta$ is an integrable distribution reducible to the submanifold $\mathcal H=0$ i.e., conditions (\ref{condiz}) hold.
$(ii)\Rightarrow(i)$. We assume that
there exist $n$ functions $R_i$ on $M$ such that the distribution $\Delta$ generated by
(\ref{gene})
is integrable and reducible on the submanifold $\mathcal H=0$, that is (\ref{gene}) satisfy (\ref{condiz}).
Hence, the first order PDE system associated with $\Delta$ (\ref{sys}) is completely integrable, that is there exist
$nl+1$ functions $f_A(q^i,c_B)$ such that $z_A=f_A(q^i,c_B)$ are a complete solution of (\ref{sys}) that is
satisfy (\ref{sys}) and the matrix $\big(\frac{\partial f_A}{\partial c_B}\big)$ has maximal rank.
The form of (\ref{sys}) implies that the solution is of separated type:
 $$
 u=\textstyle\sum_iS_i(q^i,c_B),\quad
 u_i=\textstyle\frac{d}{dq^i} S_i(q^i,c_B),\quad
\ldots \quad
u_i^{(l)}=\textstyle\frac{d^l}{(dq^i)^l} S_i(q^i,c_B)
  $$ 
Thus, $u^E=\sum_iS_i(q^i,c_B)$ is an extended solution of $\mathcal H=0$. Indeed, by construction it satisfies the completeness condition. Moreover, being the generators $D_i$ (\ref{gene}) tangent to $\mathcal H=0$ by assumption,
$\Delta$ is reducible to a distribution on $\mathcal H=0$. Then, for any $P$ in 
$\mathcal H=0$ the connected leaf $L_{(c_A)}$ passing through $P$ is entirely contained in $\mathcal H=0$.
Let us consider the submanifold of $\mathbb R^{nl+1}$ 
$\mathcal P=\{(c_B) \mid L_{(c_B)} \subset \mathcal H=0\}.$ Then, for all  $(c_B)\in \mathcal P$ 
 $u^E$ satisfies $\mathcal H=0$.
\end{proof}

\begin{proposition} \label{p:int}
In a given coordinate system $(q^i)$ on $Q$ the following statements are equivalent:
$(i)$ there exists an internal solution $u^I=\sum_iS_i(q^i,c_\alpha)$ of $\mathcal H=0$;
$(ii)$ there exist $n$ functions $R_i$ on $M$ such that the distribution $\Delta$ generated by
 the vector fields $(\ref{gene})$
is reducible to a distribution $\Delta_0$ on the submanifold $\mathcal H=0$ and $\Delta_0$ is integrable, that is 
\begin{equation} \label{condizi}
\left\{
\begin{array}{l}
D_i\mathcal H|_{\mathcal H=0}=0, \\  
D_iR_j|_{\mathcal H=0}=0, \qquad \forall \, j\neq i.
\end{array}
\right.
\end{equation}
\end{proposition}

\begin{proof}
$(i)\Rightarrow(ii)$: if there exists an internal solution, then there is also an extended solution in the same coordinates. Hence, by Proposition \ref{critE}, there are functions  $R_i$ satisfying (\ref{condiz}) and in particular
(\ref{condizi}).
$(ii)\Rightarrow(i)$:
Being $\Delta_0$ integrable, the manifold $\mathcal H=0$ is foliated by integral manifolds of $\Delta_0$ locally described
by solutions of the PDE system (\ref{sys}) depending on $nl$ parameters $(c_\alpha)$ with separated form
 $$
 u=\textstyle\sum_iS_i(q^i,c_\alpha),\quad
 u_i=\textstyle\frac{d}{dq^i} S_i(q^i,c_\alpha),\quad
\ldots \quad
u_i^{(l)}=\textstyle\frac{d^l}{(dq^i)^l} S_i(q^i,c_\alpha)
  $$ 
satisfying $\mathcal H=0$. Thus, $u^I=\sum_iS_i(q^i,c_\alpha)$ is an internal solution of $\mathcal H=0$.
\end{proof}

\begin{theorem} \label{t:3sep}
The separation of a null equation $\mathcal H=0$ occurs in a given coordinate system
if and only if one of the following equivalent conditions holds:

$1)$ there exist $n$ functions $\lambda_i(q,u,u_i,..,u_{i}^{\scriptscriptstyle{(l)}})$ such that 
 the operators $(\ref{gene})$ satisfy
\begin{equation} \label{condiz2}
\left\{
\begin{array}{l}
D_i\mathcal H=\lambda_i\mathcal H, \\  
D_iR_j=0, \qquad \forall \, j\neq i.
\end{array}
\right.
\end{equation}

$2)$ there exists (locally) a function $\Lambda(q^i,u,u_i,...,u_i^{\scriptscriptstyle{(l)}})$ such that 
the equation $\mathcal H/\Lambda=h$ is separable in the ordinary sense, i.e. the operators $(\ref{gene})$ 
satisfy
  \begin{equation} \label{sepL}
\left\{ \begin{array}{l}
D_i(\mathcal H/\Lambda)=0,\\
D_iR_j=0, \qquad \forall\, j\neq i.
\end{array}\right.
  \end{equation}

$3)$ the operators $(\ref{gene})$ satisfy
\begin{equation} \label{cond3}
\left\{ \begin{array}{l}
D_i\mathcal H =0,\\
D_iR_j|_{\mathcal H=0}=0, \qquad \forall\, j\neq i.
\end{array}\right.
  \end{equation}
\end{theorem}

\begin{proof}
Condition 1) follows straightforward from (\ref{condiz}) by applying Hadamard's Lemma: $F|_{\mathcal H=0}=0
\Leftrightarrow F=\lambda \mathcal H$.
Condition 2) is equivalent to condition 1).
Indeed, from (\ref{condiz})$_1$ we get 
$
\lambda_i=D_i\, \log \mathcal H.
$
Since $D_iD_j=D_jD_i$, we have $D_i\lambda_j=D_j \lambda_i$ and locally there exists a function $\Lambda$ such that
$$
\lambda_i=D_i\, \log \Lambda.
$$
Thus, condition (\ref{condiz})$_1$ becomes $D_i\ln \mathcal H=D_i\ln \Lambda$ and, being $D_i$ linear, 
$$
D_i(\mathcal H /\Lambda)=0.
$$
Hence, condition 1) holds iff there exists $\Lambda$ satisfying (\ref{sepL}).
Condition 3)  is a consequence of Proposition \ref{p:int}. Indeed, by applying Hadamard's Lemma to
(\ref{condizi}) and (\ref{cond3}), we get
 \begin{equation} \label{csepi}
\exists \,(\lambda_h), (\mu_{hk}) \mid 
\left\{ 
\begin{array}{l}
D_i\mathcal H=\lambda_i\mathcal H,\cr
D_iR_j=\mu_{ij}\mathcal H, \qquad \forall\, j\neq i,
\end{array}\right.
 \end{equation}
   \begin{equation} \label{csepio}
\exists \, (\nu_{hk}) \mid
\left\{ 
\begin{array}{l}
D_i\mathcal H=0,\cr \vbox{\vskip 12pt}
D_iR_j=\nu_{ij}\mathcal H, \qquad \forall\, j\neq i,
\end{array}\right.
 \end{equation}
respectively. 
We show that (\ref{csepi}) is equivalent to (\ref{csepio}).
Clearly,   (\ref{csepio}) implies (\ref{csepi}) by choosing $\lambda_i=0$.
Conversely, we
assume that (\ref{csepi}) holds and we denote by $D_i^0$, $R_i^0$ the operators (\ref{gene}) and the associated functions
$R_i$ satisfying $D_i\mathcal H=0$. moreover we denote 
 by  $D_i^\lambda$, $R_i^\lambda$ those satisfying $D_i\mathcal H=\lambda_i\mathcal H$.
Then, we have
   $$
D_i^0=D_i^\lambda-\mathcal F_i\mathcal H\frac{\partial}{\partial u_i^{\scriptscriptstyle (l)}}, \qquad
R_i^0=R^\lambda_i-\mathcal F_i\mathcal H,
  $$
where
 $
\mathcal F_i=\big(\textstyle\frac{\partial\mathcal H}{\partial u_i^{\scriptscriptstyle (l)}}\big)^{\scriptscriptstyle -1}\lambda_i ,
  $
and by a straightforward calculation we get
  $$
D_i^\lambda R_j^\lambda=D_i^0R_i^0+\mathcal H\left[\mathcal F_i \frac{\partial}{\partial u_i^{\scriptscriptstyle (l)}}\left(R_j^0+\mathcal F_j\mathcal H\right) +D_i^0\left( \mathcal F_j\right)\right].
  $$
Thus, since  $D_i^\lambda R_j^\lambda=\mu_{ij}\mathcal H$,  there exist functions $\nu_{ij}$
such that $D_i^0R_j^0=\nu_{ij}\mathcal H$.
\end{proof}

\begin{remark} \rm
The $n$ additional unknown functions $(\lambda_i)$ play the role of Lagrangian multipliers. It is remarkable the fact that the $n$ functions $(\lambda_i)$ can be replaced by a single unknown function $\Lambda$.
\end{remark}

\begin{remark} \rm
We recall that the conditions for the separation in the ordinary sense of the equation $\mathcal H=h$ are \cite{schr1,hje,KMto}
\begin{equation}\label{csepo}
\left\{ {\hbox{\hskip 0mm}}
\begin{array}{l}
D_i\mathcal H=0,\cr \vbox{\vskip 12pt}
D_iR_j=0, \qquad\qquad  (\forall\, j\neq i).
\end{array}\right.
\end{equation} 
\end{remark}


\section{\large Definition and criteria for $FER$-separation} \label{s:R-sep}
\setcounter{equation}{0}

By Proposition \ref{p:equi}, we can extend the notion of completeness for separated solutions of (\ref{nbhj})
to FER-separable solutions of (\ref{sch}) in a natural way. 
However, since $\mathcal H$ does not depend on $u$, but only on its derivatives, the solution of (\ref{nbhj}) is
defined up to an inessential additive constant (corresponding to a constant factor for $\psi$).
Hence, in this case we can consider without loss of generality  only the $2n$ functions $(z_A)=(u_i,u_{ii})$ as unknowns
(see Remark 2.4 in \cite{schr1}).
It follows that in the generators $D_i$ the term $u_i\partial_u$ can be disregarded.
Moreover, being the unknown functions $2n$, the number of parameters entering in a complete solution of the null PDE 
(\ref{nbhj}) is $2n-1$, according to Section \ref{ex2.1}.
Therefore, 
we define FER-separation as:

\begin{definition} \label{d:rsep} {\rm
The Schr\"odinger equation $(\ref{sch})$ is {\em FER-separable} in the coordinates $(q^i)$ {-- called {\em FER-separable coordinates} for $(\ref{sch})$}--
 if
there exist a value of the energy $E$ and a function $R$ such that Eq.\ $(\ref{sch})$, admits a solution of the form 
\begin{equation} \label{REsol}
\psi=R\prod_i\phi_i(q^i,c_\alpha),
\end{equation}
where $c_\alpha$ are $2n-1$ parameters satisfying the completeness condition
    \begin{equation}\label{ccond}
\mathrm{rank}\left[\begin{array}{c}
\displaystyle\frac{\partial u_i}{\partial c_\alpha} \\ \\
\displaystyle\frac{\partial v_i}{\partial c_\alpha} 
\end{array}
\right]=2n-1 ,
\qquad u_i=\displaystyle\frac{\phi'_i}{\phi_i},
\quad v_i=
\displaystyle\frac{\phi''_i}{\phi_i}.
  \end{equation}
  }
\end{definition}

\begin{remark}\label{r:cos} \rm
The meaning of completeness is the following:  
given $R$ and $E$ allowing FER-separation,
for any choice of $2n-1$ 
numbers $(c_\alpha)=(b_i,k_j)$, $i=1\dots n$, $j=1,\ldots, n-1,$ and of a point $q_0\in Q$, then
there exists a unique separated solution  $\phi$
of (\ref{nbar1}) such that 
$(\textstyle\frac{\phi'_i}{\phi_i},\textstyle\frac{\phi''_j}{\phi_j})_{q_0}=(h_i,k_j)$. 
Moreover, we observe that in Definition \ref{d:rsep} there are no assumptions on the form of the functions $\phi_i$
(as for instance $\phi_i= e^{b_iq^i}$ with $b_i\in\mathbb R$) as in the case of {\em free separation} of the 
Schr\"odinger equation \cite{schr1}. 
\end{remark}

We apply the third condition of Theorem \ref{t:3sep} to study the separation of Eq.\ (\ref{nbhj})
 that is the null equation $\mathcal H=0$ where
  \begin{equation}\label{hjv}
\mathcal H=g^{ij}u_iu_j+g^{ii}u_{ii}-\hat \Gamma^i u_i+\textstyle\frac 2{\hbar^2}E-U.
  \end{equation}
We assume here and in the following that $g^{ii}\neq 0$ for all $i$. This assumption is obviously fulfilled for a proper Riemannian manifold. This means that we are excluding null coordinates. In our case, we have $l=2$ and
$\mathcal H$ independent of $u$. Hence, both 
 the operators (\ref{gene}) and the associated functions $R_i$ do not depend on $u$:
 \begin{equation}\label{Dio}
D_i=\partial_i+u_{ii}\displaystyle\frac {\partial}{\partial u_i}+R_i\displaystyle\frac {\partial}{\partial u_{ii}}.
  \end{equation}
Since $D_i\mathcal H=0$, the functions $R_i$ have the form
  \begin{equation}\label{Rio}
R_i(q^i,u_i,u_{ii})=-\left(\frac {\partial \mathcal H}{\partial u_{ii}}\right)^{\! -1}\!\left[\partial _i\mathcal H+u_{ii}\frac {\partial \mathcal H}{\partial u_i}\right].
  \end{equation} 
We want to find conditions for which $D_iR_j|_{\mathcal H=0}=0$ ($i\neq j$).

\begin{lemma} \label{l:quad}
The function $D_iR_j$ is a polynomial of second degree in the variables $(u_{11},\ldots, u_{nn})$, whose quadratic
part  is 
\begin{equation} \label{qdp}
-2\,\frac{g^{ij}}{g^{jj}} u_{ii}u_{jj}, \qquad i,j \hbox{ n.s. }\quad i\neq j.
\end{equation}
\end{lemma}

\begin{proof}
 Recalling that 
$
\frac{\partial\mathcal H}{\partial u_{ii}}=g^{ii},
$
by inserting (\ref{hjv}) in (\ref{Rio}), we get
$$
R_j=\displaystyle\frac 1{ g^{jj}}[{}-2\, g^{jh}u_{jj}u_h-\partial _j g^{hh}u_{hh}
-\partial _j g^{hk}u_hu_k +\partial _j\hat \Gamma ^hu_h+
\hat \Gamma^ju_{jj}+\partial_jU].
$$
Since $R_j$ and all its partial derivatives contain terms at most linear in the variables $(u_{hh})$ and being 
$$
D_iR_j=\partial_iR_j+u_{ii}\frac {\partial R_j}{\partial u_i}-\frac{\partial _j g^{ii}}{ g^{jj}}R_i,
$$
a straightforward calculation shows that the only term of $D_iR_j$ which is quadratic in $(u_{hh})$ is (\ref{qdp}). 
\end{proof}

\begin{proposition} \label {p:ort}
The coordinates $(q^i)$ allowing FER-separation i.e, separation of the null equation $\mathcal H=0$ $(\ref{nbhj})$ for a suitable $R$, are necessarily orthogonal.
\end{proposition}

\begin{proof}
By condition 3) of Theorem \ref{t:3sep}, the null equation $\mathcal H=0$ is separable if and only if for all $i\neq j$, the functions
$D_iR_j$ vanish on the surface $\mathcal H=0$. 
Since the function (\ref{hjv}) is linear $(u_{hh})$, the equation  $\mathcal H=0$ is equivalent to 
$$
u_{ii}=-\frac 1{g^{ii}}\big(\sum_{\alpha\neq i}g^{\alpha\alpha}u_{\alpha \alpha} + g^{hk}u_hu_k -\hat \Gamma^h u_h+\textstyle\frac 2{\hbar^2}E-U\big).
$$
Hence, inserting the expression of $u_{ii}$  into $D_iR_j$, we get (by Lemma \ref{l:quad})
a quadratic polynomial in the $(u_{\alpha\alpha})$ $(\alpha\neq i$) with 
 quadratic part   
$$\sum_{\alpha\neq i}-2\,\displaystyle\frac{g^{ij}}{g^{jj}}\displaystyle\frac{g^{\alpha\alpha}}{g^{ii}} u_{\alpha\alpha}u_{jj},
\qquad i,j \hbox{ n.s. }\quad i\neq j,$$
which vanishes if and only if $g^{ij}=0$ for $i\neq j$.
\end{proof}

\begin{remark} \rm
By Proposition \ref{p:ort}  coordinates allowing FER-separation of 
Eq.\ (\ref{sch}) are necessarily orthogonal. 
However, nonorthogonal $R$-separation (see \cite{KMesRsep,KMnoRsep}) is possible by imposing constraints on the form of some  factors $\phi_i$: this case, which has been described in \cite{schr1} for ordinary separation of Schr\"odinger equation
and called {\em reduced separation}, is not 
examined in the present paper.
\end{remark}


In orthogonal coordinates, (\ref{hjv}) becomes
  \begin{equation}\label{hjo}
\mathcal H=g^{ii}(u_{ii}+u_i^2)-\hat \Gamma^i u_i+\textstyle\frac 2{\hbar^2}E-U
 \end{equation}
and the functions (\ref{Rio}) 
\begin{equation}\label{Rior}
R_j=\displaystyle\frac 1{ g^{jj}}[-2g^{jj}u_{jj}u_j-\partial _j g^{hh}(u_{hh}+u_h^2)
 +\partial _j\hat \Gamma ^hu_h+
\hat \Gamma^ju_{jj}+\partial_jU].
\end{equation}

In order to write in a shorter way $D_iR_j$ we introduce the following notion (\cite{hje}) 

\begin{definition}\rm \label{d:stop}
A {\em St\"ackel operator}  $S_{ij}$ is the second order differential operator  such that for any 
$f\colon Q\to \mathbb R $ 
\begin{equation} \label{stack}
 S_{ij}(f)=\partial^2_{ij}f-\partial_j\ln( g^{ii})\partial _if-\partial _i\ln( g^{jj})\partial _jf.
\end{equation}
\end{definition}


\begin{proposition} \label{c_fer}
Equation $(\ref{nbhj})$ is separable in the orthogonal coordinates $(q^i)$ 
if and only if 
  \begin{eqnarray}
&\partial _j\hat \Gamma^i-\hat \Gamma ^i\partial _j\ln  g^{ii}=0,\vphantom{\displaystyle\frac .2} \label{gam1} \\
&\displaystyle\frac {S_{ij}( g^{hh})}{ g^{hh}}-\displaystyle\frac {S_{ij}( g^{kk} )} {g^{kk}} =0, \quad \forall \;h,k,\label{csep} \\ 
& {S_{ij}(U)}\,g^{hh} - {S_{ij}(g^{hh} )}\,{(U-\textstyle\frac 2{\hbar^2}E)}=0, \quad \forall \;h. \vphantom{\displaystyle\frac 1.} \label{U1}
 \end{eqnarray}
\end{proposition}

\begin{proof}
By applying operators (\ref{Dio}) to the functions (\ref{Rior}) and using  (\ref{stack}) we get
  \begin{eqnarray}
\nonumber
& D_iR_j=\displaystyle\frac 1{ g^{jj}}
[(\partial _j\hat \Gamma^i-\hat \Gamma ^i\partial _j\ln  g^{ii})u_{ii}+
(\partial _i\hat \Gamma^j-\hat \Gamma ^j\partial _i\ln  g^{jj})u_{jj}\\
& 
-S_{ij}( g^{hh})(u_{hh}+u_h^2)+S_{ij}(\hat \Gamma^h)u_h+
S_{ij}(U)]. \label{DioRjo}
\end{eqnarray}
We impose that the function $D_iR_j$ vanishes on the surface $\mathcal H=0$ that is for
all values of $(u_h,u_h^2,u_{\alpha\alpha})$ with $\alpha \neq i$ and for
$$
u_{ii}=u_i^2-\frac 1{g^{ii}}\big(\sum_{\alpha\neq i}g^{\alpha\alpha}(u_{\alpha \alpha}+u_\alpha^2)  -\hat \Gamma^h u_h+\textstyle\frac 2{\hbar^2}E-U\big).
$$
By inserting the expression for $u_{ii}$ in (\ref{DioRjo}) we see that the only coefficient of $u_i^2$ is 
$\partial _j\hat \Gamma^i-\hat \Gamma ^i\partial _j\ln  g^{ii}$. Thus, (\ref{gam1}) must hold for any couple of 
indices $i\neq j$.
Then, assuming (\ref{gam1}), $D_iR_j|_{\mathcal H=0}$ becomes (up to a factor $g_{jj}$)
 $$ 
\bigg(\textstyle\frac{S_{ij}( g^{ii})}{g^{ii}}g^{\alpha\alpha}-S_{ij}( g^{\alpha\alpha}) \bigg)v_\alpha+
\bigg(S_{ij}(\hat \Gamma^h)-\textstyle\frac{S_{ij}( g^{ii})}{g^{ii}}\hat \Gamma^h\bigg)u_h
-\textstyle\frac{S_{ij}( g^{ii})(U-\frac 2{\hbar^2}E)}{g^{ii}}+
S_{ij}(U)=0 
$$
with $v_\alpha=u_{\alpha\alpha}+u_\alpha^2$.
By equating to zero the coefficients of $v_\alpha$, $u_h$, and of the 0-th order term, we get two conditions equivalent to (\ref{csep}), (\ref{U1}) 
and moreover
$$S_{ij}(\hat \Gamma^h)-\textstyle\frac{S_{ij}( g^{ii})}{g^{ii}}\hat \Gamma^h=0.$$
However, this last condition is disregarded being
a consequence of (\ref{gam1}--\ref{csep}), Indeed, by (\ref{gam1}) we get $S_{ij}(\hat \Gamma^h)=\hat \Gamma^h S_{ij}(g^{hh}) g_{hh}$. 
\end{proof}

\begin{remark} \rm
If $U-\textstyle\frac 2{\hbar^2}E=0$, i.e., if $R$ satisfies (\ref{sch}), then  (\ref{U1}) is always satisfied. 
\end{remark}


\begin{proposition}\label{p:sist}
If $(q^i)$ are separable coordinates for $(\ref{nbhj})$, then the function
$R$ is locally determinated by integrating  equations
  \begin{equation} \label{sist}
2\partial_i\ln R=\Gamma_i -\xi_i(q^i).
  \end{equation}
\end{proposition}

\begin{proof}
For $\hat\Gamma^i\neq 0$, we have that (\ref{gam1}) is equivalent to 
$\partial_j\big({\hat \Gamma^i}/{g^{ii}}\big)=0$ for all $j\neq i$ i.e., to $\hat\Gamma^i=g^{ii}\xi_i(q^i)$ ($i$ n.s.).
By definition of modified contracted Christoffel symbol (\ref{hatG}), since the coordinates are orthogonal, we get 
$\Gamma_i-2\partial_i\ln R=\xi_i(q^i)$, that is (\ref{sist}). 
The differential system is integrable
in orthogonal coordinates satisfying $(\ref{csep})$. Indeed, a straightforward calculation shows that  
 $$
 \partial _j \Gamma_i=\partial _i \Gamma_j \Leftrightarrow \partial^2_{ij} \ln g^{jj}=\partial^2_{ij} \ln g^{ii}
 \Leftrightarrow \textstyle\frac{S_{ij}(g^{jj})}{g^{jj}}=\textstyle\frac{S_{ij}(g^{ii})}{g^{ii}}.$$
\end{proof}

\begin{remark} \rm
By using (\ref{sist}) we can write $\Delta R/R$ (included in the modified potential $U$) in terms of the contracted Christoffel symbols $\Gamma_i$ as
\begin{equation} \label{E-U}
\displaystyle\frac {\Delta R}R=\displaystyle\frac 14 g^{ii}\big(2\partial_i\Gamma_i-\Gamma_i^2+\xi_i^2-2\partial_i\xi_i\big).
\end{equation}
\end{remark}


We shortly recall some definitions arising from the theory of separation of geodesic Hamilton-Jacobi equation
 (see \cite{hje}).
\begin{definition} \rm
Orthogonal coordinates  $(q^i)$ are {\em separable} if 
the metric $g^{ii}$ is a {\em St\"ackel metric} i.e.  it is a  row of the inverse of a regular matrix 
$S=(\varphi^{\scriptscriptstyle{(j)}}_i)$ ({\em St\"ackel matrix})
whose elements depend on the coordinate corresponding to the lower index only. 
Orthogonal coordinates $(q^i)$ are {\em conformal separable} if
the metric $g^{ii}$ is a {\em conformal St\"ackel metric} i.e., 
there exists a function $\Lambda$ such that $g^{ii}/\Lambda$ is a St\"ackel metric.
A function $f$ is  a {\em pseudo-St\"ackel factor} (resp. {\em St\"ackel factor}),
if it can be written as $f=g^{ii}\phi_i(q^i)$, where $(g^{ii})$ is a conformal  St\"ackel metric (resp.   St\"ackel metric).
\end{definition}

\begin{theorem} \label{t:res}
Necessary and sufficient conditions for the FER-separation of $(\ref{sch})$ in a given coordinate system $(q^i)$ (under the assumption $g^{ii}\neq 0\; \forall i$) are: \\
$(1)$ 
the coordinates are orthogonal;
$(2)$ 
the coordinates are conformal separable;
$(3)$ 
the function $$\textstyle\frac 2{\hbar^2}(E-V)+\textstyle\frac 14 g^{ii}(2\partial_i\Gamma_i-\Gamma_i^2)$$ is a 
pseudo-St\"ackel factor.
In this case $R$ is any solution of equations $(\ref{sist})$.
\end{theorem}

\begin{proof} 
By Proposition \ref{p:ort}, we have that coordinates are necessarily orthogonal. Moreover, by Proposition \ref{c_fer}
in orthogonal coordinates FER-separation occurs if and only if conditions
(\ref{gam1}), (\ref{csep}) and (\ref{U1}) hold.
Equations (\ref{csep})  means that the orthogonal coordinates $q^i$ are conformal separable coordinates, condition $(\ref{U1})$ means that $U-\textstyle\frac 2{\hbar^2}E$ is a  pseudo-St\"ackel factor (see \cite{hje}) and, by Proposition \ref{p:sist}, Eq.\ (\ref{sist}) determines $R$ up to separated factors. 
By (\ref{E-U}), we have that we can replace $U-\textstyle\frac 2{\hbar^2}E$ by the function $\frac 2{\hbar^2}(E-V)+\frac 14 g^{ii}(\partial_i\Gamma_i-\frac 12 \Gamma_i^2)$, disregarding the term $\frac 18 g^{ii}\theta_i$ which is a pseudo-St\"ackel factor for any choice of $\xi_i$ in (\ref{sist}).
\end{proof}


\begin{remark} \rm
In the ordinary separation for Schr\"odinger equation, condition (3) is split into the 
 compatibility condition for the potential $(V=g^{ii}\theta_i(q^i))$ and the Robertson condition $(\partial_i\Gamma_j=0)$. 
Moreover, the coordinates are in this case necessarily separable, while in the case of FER-separation conformal separable coordinates are also allowed.
\end{remark}

\begin{remark} \rm
For the Laplace equation, condition (3) of Theorem \ref{t:res} means that 
$\frac{1}4 g^{hh}(2\partial_h\Gamma_h-\Gamma_h^2)$ is a pseudo-St\"ackel factor, that is 
\begin{equation} \label{rrob}
\exists \, \theta_i(q^i)\mid
\displaystyle\frac{g^{hh}}4 (2\partial_h\Gamma_h-\Gamma_h^2)=g^{ii}\theta_i(q^i).
\end{equation}
We remark that this condition in general does not imply  (see \cite{schr2}),
\begin{equation}\label{prer}
2\,\partial_i\Gamma_i-\Gamma_i^2=4\theta_i(q^i). \quad \forall\;i
\end{equation}
For example toroidal coordinates (see Section \ref{s:es}) satisfy (\ref{rrob}) since
$g^{hh}(2\partial_h\Gamma_h-\Gamma_h^2)=g^{33},$ 
but they do not satisfy (\ref{prer}). Moreover, even if the $(q^i)$ are separable, (\ref{rrob}) does not imply the Robertson condition 
$\partial_i\Gamma_j=0$ (by (\ref{sist}) Robertson condition implies that $R$ is a separated function and $R$-separation is then called {\em trivial} \cite{KMto}).
A counterexample of that
is given by the conformally flat metric allowing non trivial $R$-separation in \cite{KMesRsep}. In \cite{KMoRsep}   condition (\ref{prer}) is linked to the fact that symmetry operators can be put
in the reduced form. 
\end{remark}

\begin{proposition}  If $(q^i)$ are FER-separable coordinates for $(\ref{sch})$ for two values of the energy $E_1\neq E_2$, then $(q^i)$ are separable and allow $R$-separation for all values of the energy.
\end{proposition}
\begin{proof}
If (\ref{U1}) holds for both $E_1$ and $E_2$, then $S_{ij}(g^{kk})=0$, $k=1\dots n$ and (\ref{U1}) hold for all $E\in \mathbb R$.
\end{proof}

\subsection{Separated equations }

If $q^i$ are conformal separable coordinates for the orthogonal metric $\mathbf G =(g^{ii})$, then 
for any pseudo-St\"ackel factor $f$ we have that $(g^{ii}/f)$ is a St\"ackel metric. 
Hence, in coordinates allowing FER-separation we have that the conformal metric $\bar {\mathbf G}$ with components
\begin{equation} \label{gbar}
\bar g^{jj}=\frac{{g^{jj}}}{\frac 2{\hbar^2}E-U}
\end{equation}
 is a St\"ackel metric and that Eq.\ (\ref{nbhj}) is equivalent
to 
\begin{equation} \label{eqsep0}
\bar g^{ii}(u_{ii}+u_i^2-\xi_iu_i)=-1,
\end{equation}
for any choice of the functions $\xi_i(q^i)$. 
In order to integrate Eq.\ (\ref{eqsep0})by separation of variables, it should be considered as the $n$-th equation 
of the system
  $$
\varphi_{(j)}^i(u_{ii}+u_i^2-\xi_iu_i)=a_j, \qquad (a_n=-1)
  $$
where $\varphi_{(j)}^i$ is the inverse of a St\"ackel matrix
associated with the St\"ackel metric $\bar{\mathbf G}$ and $(a_1,\ldots,a_n)$ are real constants.
By inverting this system, we get the separated equations
  \begin{equation}\label{seq1}
u_{ii}+u_i^2-\xi_i u_i=\sum_{\alpha=1}^{n-1}a_\alpha\varphi^{(\alpha)}_i-\varphi^{(n)}_i,
\end{equation}
corresponding by using (\ref{der}) to the Riccati equations for the separated factors $\phi_i$
\begin{equation}\label{seq2}
\phi_i''-\xi_i\,\phi_i'-\varphi_i^{\scriptscriptstyle{(j)}}\,a_j\,\phi_i=0,\qquad (a_n=-1).
\end{equation}

Being $R$ locally determined (up to separated factors) by Eq.\ (\ref{sist}), it is evident that we always can choose $\xi_i=0$ for $i=1, \dots , n$ so that (\ref{sist}) becomes
\begin{equation}\label{rcan}
2\partial _i\ln R=\Gamma_i.
\end{equation}
It follows that

\begin{proposition}
In FER-separable  coordinates, by choosing $R$ satisfying $(\ref{rcan})$, separated equations $(\ref{seq1})$, $(\ref{seq2})$  have the canonical form
$$
u_{ii}+u_i^2=a_j\varphi^{(j)}_i,
$$
$$
\phi_i''=\varphi_i^{\scriptscriptstyle{(j)}}\,a_j\,\phi_i.
$$
\end{proposition}

\begin{remark} \rm
Even if they  do not explicitly appear in the separated equations, the potential $V$ and the value $E$ are contained in the components St\"ackel metric $[\varphi_i^{\scriptscriptstyle{(j)}}]$
\end{remark}

\begin{remark} \rm
The above described method of separating the variables gives rise to two kinds of constants in the separated
solutions: the $n-1$ constants $a_j$ ($j=1,\ldots,n-1)$ of Eq.\ (\ref{seq2}) that we call separation constants and the integration constants of Eq.\ (\ref{seq2}). Even if the 2nd order Eq. (\ref{seq2}) give rise to $2n$ constants,  $n$ of them are inessential constant factors of the $\phi_i$, irrelevant for our purposes as explained at the beginning of the Section. 
Therefore, there are only $n$ non-trivial integration constants ($b_i= (\phi_i'/\phi_i)_{q_0}$) which, together with the $n-1$ $(a_j)$, form
the set of $2n-1$ constants satisfying the completeness condition (\ref{ccond}).
The relation between the constants $(a_j,b_i)$ and the constants ($b_i,k_j$)  introduced in Remark \ref{r:cos} 
with $(k_j)= (\phi_i''/\phi_i)_{q_0})$ is given by 
 (\ref{seq2}) evalueted in the initial point $q_0$.
 \end{remark}
 
 In the following subsections, we shall see that, in analogy with the ordinary separation of the Schr\"odinger equation (see \cite{KMto, schr1, schr2} and references 
therein), the separation constants are in fact the eigenvalues of
$n$ second-order pairwise-commuting conformal symmetry operators of the equation related to quadratic first-integrals in involution of an associated Hamilton-Jacobi equation.

\subsection{Conformal Killing Tensors and FER-separation}
Fixed-Energy separation of (\ref{nbhj}) is characterized by the existence of CKTs of the metric $\mathbf G$ in conformal involution with common eigenvectors (see \cite{hje}).
Indeed,  conformal separable coordinates are associated with  $n$ symmetric 2-tensors $\mathbf K_1,\ldots
\mathbf K_n=\bar{\mathbf G}$ which are Killing tensors with respect to the metric $\bar{\mathbf G}$ (\ref{gbar}), 
 simultaneously diagonalized in the $(q^i)$, and with contravariant components
\begin{equation} \label{comp}
K^{ii}_j=\varphi^i_{(j)}, \qquad K_j^{ih}=0, \quad \hbox{for $i\neq h.$}
\end{equation}
Let us consider the cotangent bundle $T^*Q$ with canonical coordinates $(q^i,p_i)$ and Poisson brackets of functions $A$, $B$ on $T^*Q$ defined by
$$
\{A,B\}=\Sigma_{i=1}^n\displaystyle\frac {\partial A}{\partial p_i}\displaystyle\frac {\partial B}{\partial q^i}-\displaystyle\frac {\partial A}{\partial q^i}\displaystyle\frac {\partial B}{\partial p_i}.
$$
Let us construct the quadratic polynomials in the momenta associated with ($\mathbf K_i$)

\begin{equation} \label{pol}
P_{K_h}=K_h^{ij}p_ip_j \qquad h\neq n, \qquad P_{K_n}=\bar g^{ij}p_ip_j.
\end{equation}
We have that \cite{KMckt, hje} the functions $P_{K_h}$ defined by $(\ref{pol})$ and $(\ref{comp})$ satisfy
for $h$, $j=1, \dots ,n$: 
\begin{equation} \label{invo}
\{P_{K_h},P_{K_j}\}=0,
\end{equation}
i.e.,  $P_{K_h}$ are pairwise in involution first integrals of the geodesic Hamiltonian $P_{K_n}$, 
and the geodesic Hamilton-Jacobi equation
$
\bar g^{ij}p_ip_j=h
$
is separable in the $(q^i)$.
Hence,  we get 
\begin{proposition}\label{p:intr}
An intrinsic necessary condition for the existence of FER-separable coordinates for the Schr\"odinger equation 
$(\ref{sch})$ is that there exist $n-1$ CKTs $\mathbf K_1,\ldots
\mathbf K_{n-1}$ such that  $(i)$ $\mathbf K_1,\ldots
\mathbf K_{n-1},{\mathbf G}$ are linearly independent, $(ii)$ $\mathbf K_1,\ldots
\mathbf K_{n-1}$ have common eigenvectors, and $(iii)$ Eq. $(\ref{invo})$ is satisfied. 
\end{proposition}



\subsection{Symmetry operators}
Starting from  separated equations (\ref{seq1}) and (\ref{seq2}) we build conformal symmetry operators of (\ref{sch}). Interpreting the left-hand side  of (\ref{seq1}) as  the results of $n$ differential operators acting on $u$,
\begin{equation} \label{ope0}
u\mapsto\varphi_{(j)}^i(u_{ii}+u_i^2-\xi_iu_i),
\end{equation}
and inserting (\ref{der}), (\ref{sist}), (\ref{rsol}) in (\ref{ope0}), we get $n$ operators $H_j$ on $\psi$ defined by
\begin{equation}\label{ope1}
H_j\psi=\Delta_j\psi-\displaystyle\frac 1R\Delta_jR\psi,
\end{equation}
where
\begin{equation}\label{ope2}
\Delta_j\colon \psi\mapsto \varphi^i_{(j)}(\partial_{ii}^2\psi-\Gamma_i\partial_i\psi)=\varphi^i_{(j)}\delta_i\psi, \qquad \delta_i\colon \psi\mapsto\partial_{ii}^2\psi-\Gamma_i\partial_i\psi.
\end{equation}
For $j=n$ we have $\Delta_n=\Delta/(\frac 2{\hbar^2}E-U)$ and for $a_n=-1$, $H_n\psi=a_n\psi$ is equivalent to the Schr\"odinger equation (\ref{sch}).
By Proposition \ref{p:sist}, since  $\varphi_{(j)}^i /\bar g^{ii}$ are  
the eigenvalues $\lambda^i_{j}$ w.r. to $\bar{\mathbf G}$ of $n$ diagonalized KTs of $\bar{\mathbf G}$, defined by
$$
K^{ii}_{(j)}=\varphi_{(j)}^i =\lambda^i_{j}\bar g^{ii},
$$
and since the $\lambda_j^h$ satisfy the following intrinsic Killing-Eisenhart equations (see \cite{eis,KMckt,hje})
$$
\partial_i\lambda_j^h=(\lambda_j^i-\lambda_j^h)\partial_i\log \bar g^{hh},
$$
a direct computation shows that

\begin{proposition}
Let $(q^i)$ be orthogonal conformal separable coordinates and $(\mathbf K_j)$ be $n$ independent KT of the conformal metric $\bar{\mathbf G}$ (\ref{gbar}) in involution and simultaneously diagonalized $(K_j^{ih}=\delta^{ih}\varphi_{j}^i)$, with $\mathbf K_n=\bar {\mathbf G}$. Then, 
the operators $H_i$ pairwise commute i.e., for $j,k=1,\dots, n$
\begin{equation}
[H_j,H_k]=H_jH_k-H_kH_j=0.
\end{equation}
\end{proposition}

By expanding $[H_i,S]$, where $S$ denotes the Sch\"odinger operator
$S:\psi\mapsto -\Delta\psi+\textstyle\frac {2}{\hbar^2}V$, by (\ref{ope1}) and (\ref{ope2}), we see that $H_i$ are conformal symmetry operators \cite{KMLapl} for $S$. It follows that

\begin{proposition}
FER-separable solutions $\psi$ are common eigenvectors of commuting second order conformal symmetry operators
for the Sch\"odinger operator
$S$. 
\end{proposition}


\section{An example: toroidal coordinates on the Euclidean three-space} \label{s:es}
\setcounter{equation}{0}

Let us consider toroidal coordinates $(q^i)=(\eta, \theta,\varphi)$ on the Euclidean three-space $\mathbb E_3$.
These coordinates have been applied also to biophysical systems \cite{biol}.
The transformations to Cartesian coordinates $(x,y,z)$ are
$$
\left\{
\begin{array}{l}
x=\frac{a\sinh\eta \cos\varphi}{\cosh \eta-\cos\theta},\\
y=\frac{a\sinh\eta\sin\varphi}{\cosh \eta-\cos\theta},\\
z=\frac{a\sin\theta}{\cosh \eta-\cos\theta}.\\
\end{array}
\right.
\qquad a\in \mathbb R^+
$$
The coordinates hypersurfaces $q^i=\mathrm{const.}$ are toroids, spherical bowls, and half-planes through the $z$-axis, respectively \cite{MSb}. The non vanishing contravariant components of the Euclidean metric in the $(q^i)$ are
$$
g^{11}=g^{22}=\frac{(\cosh \eta-\cos\theta)^2}{a^2},\qquad g^{33}=\frac{(\cosh \eta-\cos\theta)^2}{a^2\sinh^2 \eta}.
$$
We consider the FER-separation of Schr\"odinger equation for $E=0$
\begin{equation} \label{sch0}
\Delta \psi -\textstyle\frac {2}{\hbar^2}V\psi=0.
\end{equation}
Since for toroidal coordinates we have $g^{hh}(2\partial_h\Gamma_h-\Gamma_h^2)=g^{33},$ which is a St\"ackel factor,
condition (3) of Theorem \ref{t:res} provides the form of all possible potentials allowing FER-separation of (\ref{sch0})
$$
V=(\cosh \eta-\cos\theta)^2\left(f_1(\eta)+f_2(\theta)+\frac 1{\sinh^2 \eta}f_3(\varphi)\right),
$$
where $f_i$ are arbitrary functions.
By integrating (\ref{sist}) we get 
\begin{equation} \label{can}
R=\bigg(\frac{\cosh \eta-\cos\theta}{\sinh\eta}\bigg)^{\frac12}
\end{equation} for all $\xi_i=0$ (canonical separated equations), while we get $R=({\cosh \eta-\cos\theta})^{\frac12}$ for the choice $\xi_1=\mathrm{cotanh}\eta$
(non canonical separated equations))  
The conformal metric $\bar g^{ii}=g^{ii}/\sigma$ with
$$
\sigma=(\cosh \eta-\cos\theta)^2\left(f_1(\eta)+f_2(\theta)+{(\sinh\eta)}^{-2}f_3(\varphi)\right)
$$
is a St\"ackel metric and associated with the St\"ackel matrix (depending on $V$)
$$
S=\left[
\begin{array}{ccc}
f_1 & -1 &-\sinh^{-2}\eta\\
f_2 & 1 & 0 \\
f_3 & 0 & 1 \\
\end{array}
\right].
$$
The separated equations are
\begin{eqnarray*}
\frac{d^2\phi_1}{d\eta^2}+\mathrm{cotanh} \eta \frac{d\phi_1}{d\eta}+(f_1-c_2-&c_3(\sinh\eta)^{-2})\phi_1=0 \\
\frac{d^2\phi_2}{d\theta^2} +(f_2+c_2)\phi_2=0& \\
\frac{d^2\phi_3}{d\varphi^2} +(f_3+c_3)\phi_3=0 &\\
\end{eqnarray*}
where the term in $\frac{d\phi_1}{d\eta}$ disappears if $R$ is chosen to be (\ref{can}). Two conformal Killing tensors in involution associated with toroidal coordinates are
$$
\mathbf K_1=\partial_\theta \otimes\partial_\theta +f_2 \bar{\mathbf G}\qquad
\mathbf K_2=\partial_\varphi \otimes\partial_\varphi+ f_3 \bar{\mathbf G}
$$
whose components in Cartesian coordinates (disregarding the term proportional to the metric tensor) are
$$
\mathbf K_1=\frac{1}{2a^2}\left[
\begin{array}{ccc}
2x^2z^2 & 2xyz^2 & -(x^2+y^2+z^2)xz\\
'' & 2z^2y^2 & -(x^2+y^2+z^2)yz \\
'' & '' & (x^2+y^2+z^2)^2 \\
\end{array}
\right]
\qquad
\mathbf K_2=\left[
\begin{array}{ccc}
\;y^2 & -xy & 0\\
 -xy& \; x^2 & 0 \\
 0& 0 &  0 \\
\end{array}
\right].
$$
The research of non trivial examples of FER-separation for a single value $E\neq 0$ in dimension $n>2$ is in progress.  

\section*{Acknowledgments}
The authors wish to thanks G. Marmo, R.McLenaghan, R. Smirnov and M. Chanachowicz for useful discussions on these topics.
This research is partially supported by MIUR and by Lagrange Project of Fondazione CRT.



\begin{thebibliography}{0}
\bibitem{BeSepCo} S.\ Benenti, Connections and Hamiltonian Mechanics, in {\it Gravitation Electromagnetism
and Geometrical Structures}, ed. G.Ferrarese  (Pitagora Eds.\ 
Bologna 1996) pp.~185-206.

\bibitem{schr1} S.\ Benenti, C.\ Chanu, and G.\ Rastelli, Remarks on the connection between the
additive separation of the Hamilton-Jacobi equation
and the multiplicative separation of the 
Schr\"odinger equations. I. The completeness and Robertson conditions, {\it J. Math. Phys.} {\bf 43} (2002), 
5183--5222. 

\bibitem{schr2} S.\ Benenti, C.\ Chanu, and G.\ Rastelli, 
Remarks on the connection between the
additive separation of the Hamilton-Jacobi equation
and the multiplicative separation of the 
Schr\"odinger equations. II. First integrals and symmetries operators,
{\it J. Math. Phys.} {\bf 43} (2002), 5223--5253.

\bibitem{hje} S.\ Benenti, C.\ Chanu, and G.\ Rastelli, Variable separation theory for the null Hamilton--Jacobi equation, {\it J. Math. Phys.} {\bf 46} (2005), 042901/29.

\bibitem{Marmo} A.\ D'Avanzo, G. Marmo and A.\ Valentino, Reduction and Unfolding for Quantum Systems: the Hydrogen Atom,  math-ph/0504033.

\bibitem{eis} L.P.\ Eisenhart, {\it Riemannian geometry},  (Princeton univ. Press, Princeton, 1949).

\bibitem{KM4dR}  E.G.\ Kalnins and W. Miller Jr., $R$-separation of variables for the four-dimensional flat space Laplace and Hamilton-Jacobi equations {\it Trans. Amer. Math. Soc.} {\bf 244} (1978), 241--261. 

\bibitem{KMesRsep}  E.G.\ Kalnins and W. Miller Jr., . Some remarkable $R$-separable coordinate systems for the Helmholtz equation, {\it Lett. Math. Phys.} {\bf 4} (1980), 1047--1053.

\bibitem{KMLapl} E.G.\ Kalnins and W. Miller Jr., Intrinsic characterisation of orthogonal $R$-separation for Laplace equation {\it J.\ Phys. A} {\bf 15} (1982) 2699--2709 

\bibitem{KMto} E.G.\ Kalnins and W. Miller Jr., Intrinsic characterization of variable separation for the partial 
differential equations of Mechanics, in Proceedings of IUTAM-ISIMM Symposium on {\it Modern Developments
in Analytical Mechanics}, Torino 1982, Atti Accad.\ Sci.\ Torino 
{\bf 117}, Vol.2, (1983) pp.~511--533.

\bibitem{KMnoRsep}  E.G.\ Kalnins and W. Miller Jr., The general theory of $R$-separation for the Helmholtz equation, {\it J. Math. Phys.} {\bf 24} (1983), 1047--1053.

\bibitem{KMckt} E.G.\ Kalnins and W. Miller Jr., Conformal Killing tensors and variable separation for Hamilton-Jacobi equations,  {\it SIAM J.\ Math Anal.}, {\bf 14} (1983) 126--137.

\bibitem{KMoRsep} E.G.\ Kalnins and W. Miller Jr., The theory of orthogonal $R$-separation for Helmholtz
equation {\it Adv.\ in Math.} {\bf 51} (1984), 91--106.

\bibitem{biol} S.\ Kuyucak, M.\ Hoyles, and S.H.\ Chung, Analytical Solutions of Poisson's Equation for Realistic Geometrical Shapes of Membrane Ion Channels, {\it Biophys J.} {\bf 74} (1998),  22--36.

\bibitem{MS52} P.\ Moon and D.E.\ Spencer, Separability conditions for the laplace and Helmholtz equations, {\it J.\ Franklin Inst.} {\bf 253} (1952), 585--600.

\bibitem{MS52b} P.\ Moon and D.E.\ Spencer, Theorems on separability in riemannian $n$-space, {\it Proc.\ Amer.\ Math.\ soc.} {\bf 3} (1952), 635--642.

\bibitem{MSb} P.\ Moon and D.E.\ Spencer {\it Field theory handbook}, (Springer, Berlin, 1961).

\end{thebibliography}
\end{document}